\documentclass[article]{JHEP3} 

\usepackage{epsfig,multicol,bbm}
\newcommand\fverb{\setbox\fverbbox=\hbox\bgroup\verb}
\newcommand\fverbdo{\egroup\medskip\noindent%
			\fbox{\unhbox\fverbbox}\ }
\newcommand\fverbit{\egroup\item[\fbox{\unhbox\fverbbox}]}
\newbox\fverbbox

\title{Holographic entanglement entropy in imbalanced superconductors}

\author{Arghya Dutta$^a$, Sujoy Kumar Modak$^{b,c}$\\
	$^a$S. N. Bose National Centre for Basic Sciences,\\ Block-JD, Sector-III, Salt Lake City, Kolkata - 700098, India. \\\\
	$^b$IUCAA, Post Bag 4, Ganeshkhind, Pune University Campus, Pune - 411 007, India.\\
  $^c$Instituto de Ciencias Nucleares, Universidad Nacional Aut\'{o}noma de M\'{e}xico,\\ Apartado Postal 70-543, Distrito Federal, 04510, M\'{e}xico.\\\\
    E-mail: \email{arghya@bose.res.in}\\
    E-mail: \email{sujoy.kumar@correo.nucleares.unam.mx}}

\abstract{We study the behavior of holographic entanglement entropy (HEE) for imbalanced holographic superconductors. 
We employ a numerical approach to consider the robust case of fully back-reacted gravity system. The hairy black hole solution is found by using our numerical scheme. Then it is used to compute the HEE for the superconducting case. The cases we study show that in presence of a mismatch between two chemical potentials,
below the critical temperature, superconducting phase has a lower HEE in comparison to the AdS-Reissner-Nordstr\"{o}m black hole phase. Interestingly, the effects of chemical imbalance are different in the contexts of black hole and superconducting phases. For black hole, HEE increases with increasing imbalance parameter while it behaves oppositely for the superconducting phase. The implications of these results are discussed.}

\keywords{Holography and condensed matter physics (AdS/CMT), Gauge-gravity correspondence, AdS-CFT Correspondence}

\begin{document} 


\section{Introduction}
Holography is a remarkable concept that plays vital role to understand many features in modern physics-- starting from black holes and cosmology to AdS/CFT correspondence. Historically it was first realized through the expression of black hole entropy \cite{bek,hawk} 
\begin{equation}
S_{BH}=\frac{{\textrm{Area}}(\Sigma_H)}{4 G_N}
\label{BHE}
\end{equation}
which was found surprisingly proportional to the horizon area and not the volume. It motivates one to think that the bulk degrees of freedom somehow ``holographically" mapped to the surface/horizon degrees of freedom which results this non-extensive behavior in entropy. Later on this enabled 't Hooft, Susskind and others \cite{hooft}-\cite{susk3} to explain our Universe using the concept of holography. Most recent additions to this list are AdS/CFT correspondence and entanglement entropy.

AdS/CFT correspondence, first conjectured by Maldacena \cite{Maldacena:1997re},  is a realization of much discussed proposition of 't Hooft \cite{hooft2} on the large $N$ limit of strong interactions. AdS/CFT correspondence states that a supergravity theory in $AdS_5\times S^{5}$ is a ``dual" description of strongly coupled ${\cal N}=4,~SU(N)$ SYM theory ``residing" in its boundary in the limit of $N\rightarrow\infty$. Here $S^5$ is compactified to a radius $L>>l_s$ ($l_s$= string length) which is also the radius of curvature of $AdS$ spacetime. Therefore effectively a five dimensional gravity theory is ``holographically" reduced to a four dimensional conformal field theory. This ``duality" in two theories was quantified by Witten \cite{Witten:1998qj},  by identifying the bulk field with boundary operator and $n$ point correlation functions in terms of derivatives of the gravitational partition function with respect to the boundary value of that field. In support of this yet unproven AdS-CFT correspondence, 
there exists many direct and indirect evidences, for example--(i) the isometry group $SO(4,2)$ of $AdS_5$ is isomorphic to the conformal group of the SYM theory, (ii) matching of correlation functions calculated separately from CFT and that using AdS/CFT tool and many others (for more see reviews \cite{Aharony:1999ti}-\cite{D'Hoker:2002aw}), which make it robust. It is true that the exact reason/s why such two apparently different theories should behave so cohesively is/are not known, but the role of holography is undeniable, and therefore it needs further attention. The major applications of this correspondence can be broadly classified in two parts: one which are in the context of QCD (for a review see \cite{erdrev}) and the other in the context of condensed matter physics \cite{Hartnoll:2008vx, Hartnoll:2009sz}. It is the second case which is our interest in this paper.

The role of holography in the much focussed issue of entanglement entropy has been recently highlighted by Ryu and Takayanagi \cite{Ryu:2006bv, Ryu:2006ef}. If a system, described by certain quantum field theory or some quantum many body theory, is divided into two parts, say $A$ and $B$, then entanglement entropy $S_A$ of the subsystem $A$ is a non-local quantity which measures how the above systems are correlated, quantum mechanically, with each other. In defining $S_A$ one traces out the degrees of freedom of the space-like submanifold $B$ which is not accessible to an observer in $A$. Anyone familiar with the concept of black hole entropy would find this definition very much analogous to the case where an observer outside the black hole event horizon has no access to the information inside. Indeed this is one of the motivation for the authors of \cite{Ryu:2006bv, Ryu:2006ef} to {\it heuristically} propose an ``holographic" formula of entanglement entropy, given by
\begin{eqnarray}
S_A= \frac{{\textrm{Area}}(\gamma_A)}{4 G_{N}}
\label{HEE}
\end{eqnarray}
where $\gamma_A$ is the $d$ dimensional surface whose $d-1$ dimensional boundary $\partial_{\gamma_{A}}$ matches with the boundary $\partial_A$ of the field theory subsystem $A$ ({\it see} Figure 1). Of course the choice for such a surface is not unique. In this context it is suggested that this surface, among various choices, should be the {\it minimal}. This minimal surface is found by extremizing the area functional and finding out the solution (in case there are more than one) whose area takes the minimum value. 
\begin{figure}
\centering
\includegraphics[scale=0.6]{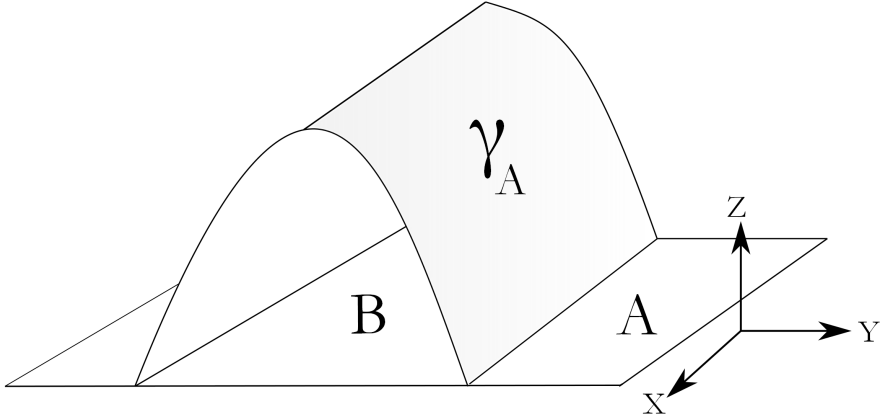}
\caption{Schematic diagram of the computational scheme of holographic entanglement entropy via AdS/CFT. The field theory system ``resides'' on the planar portion, whereas, the minimal (entangled) surface $\gamma_A$ is extended inside the bulk ($z$ axis) towards the horizon (not shown in the figure).}
\label{eeadsrn}
\end{figure}

At the present status the HEE formula (\ref{HEE}) is not conclusively proven{\footnote{Refer to \cite{Fursaev:2006ih} for an attempt and others \cite{Headrick:2010zt, casini} for more details.}}. Nevertheless there is a list of evidences which bolsters the robustness of this formula. One direct evidence comes from the $\textrm{AdS}_{3}/\textrm{CFT}_{2}$ context where the CFT result of the entanglement entropy $S_{A}=\frac{c}{3}\log{\frac{l}{a}}$, matches with the holographic calculation, in which $l$ is the width of the subsystem $A$ and $c=\frac{3R}{2G_{N}}$ relates the central charge $c$ with the radius of curvature $R$ of the $AdS_3$ spacetime. Although this evidence has not been explicitly seen in higher dimensional cases ($AdS_{d+1}/CFT_{d}$ with $d>2$), there are more compelling arguments which put confidence on (\ref{HEE})  ({ \it for details see reviews} \cite{Nishioka:2009un, Takayanagi:2012kg} {\it  and references therein}). The major usefulness of the HEE is the same as the basic principle of AdS/
CFT: overcoming the computational difficulties of complex many body field theoretic calculations in terms of much more simpler classical gravity calculations.   

Our work in this paper is motivated by a recent study by Albash and Johnson \cite{Albash:2012pd, Albash:2010mv}, where it is argued that HEE might be an useful physical quantity for characterizing holographic superconductors. They found that the finite part of the HEE ($S_{f}$) of superconducting and non-superconducting phases follow a pattern which enables one to identify the phase of a system. For a given system size and for all temperatures below the critical value $T_c$, $S_f$ takes a lower value for the superconducting phase compared to its value for the corresponding non-superconducting (black hole) phase. Whereas for temperature higher than $T_c$, where no superconducting state appears, $S_f$ only exist for the latter phase. The reason behind the smaller value of HEE for the superconducting state is explained in terms of number of the degrees of freedom that the system possesses. This number is higher in the black hole phase but as the superconductor forms some of them are condensed and results into a lower HEE.  Further works in this direction are also reported in \cite{Cai:2012nm}-\cite{Cai:2013oma}. It should also be mentioned that apart from the finite value of HEE given by $S_{f}$ there is also a diverging part. However, such a divergence is not the characteristic of the holographic calculation only, it also appears in the continuum limit of the conformal field theory calculations. One can avoid such diverging terms by introducing a UV cut off through the introduction of a lattice spacing in the expression of entanglement entropy. In the holographic calculation, the divergences can be avoided if the boundary of the minimal surface is chosen slightly away from the asymptotic infinity by choosing the appropriate limit of the radial coordinate.  

In this paper we explore the behavior of HEE in an {\it imbalanced mixture} of two fermionic systems with a mismatch in their chemical potentials \cite{Bigazzi11}. One motivation of choosing the imbalanced system is that these are quite interesting in the condensed matter framework. This is discussed in more detail in section (2.1). Our aim is to compute the HEE for two phases (black hole and superconducting phases) and compare their numerical values as a function of the strip size. The gravitational system is considered to have the backreaction term. We use numerical shooting method to find the hairy black hole solution for two different values of chemical mismatch $0.01$ and $0.02$ and compare the results for HEE with the black hole phase with same chemical potential and temperature. For both cases we find that HEE for the superconducting phase stay below the black hole phase. On the other hand the effect of the imbalance on HEE is exactly opposite for the superconducting phase than the black hole phase. While HEE increases with the increase in imbalance parameter for black holes it decreases when superconducting state forms. This demands a more careful interpretation whether HEE can always be used for identifying the preferable state or not. The reason being one expects that for increasing chemical imbalance the superconducting state is hard to achieve and at a certain larger value this state disappears. But from our study it appears that the gap of HEE between the black hole and superconductor only increases. Superconducting state is more and more stable with the increase in chemical imbalance if one solely rely on HEE. And in that way one never gets rid of the superconducting state. This contradicts the usual expectations.  

We organize this paper in the following manner. In the next section we set the platform by introducing the imbalanced system from condensed matter and holographic perspectives. Section 3 is devoted for providing the equations of motion whose solutions are discussed in next two sections (4, 5). In Section 4 we consider the case where only RN-AdS black hole solution exists and compute its HEE for various values of the imbalanced parameter. In Section 5 
we consider the case where superconducting state appears. For that we make use of numerical method and compute the hairy black hole metric. Then this metric is used to compute the HEE for different chemical mismatches. In both cases we plot HEE with respect to the system strip width $l$ of the field theory subsystem and compare the values of HEE for various cases. Finally we conclude in Section 6.

\section{Imbalanced superconducting systems}
\subsection{Condensed matter description}
Imbalance in the population of spin-up and spin-down fermions leads to exotic superconducting states. In the context of solid-state superconductors the existence of these exotic superconducting states were theoretically proposed in 1960s by Sarma\cite{sarma} and Maki\cite{maki} in high magnetic field and low temperatures. Soon after Fulde and Ferrell\cite{ff} and Larkin and Ovchhinikov\cite{lo} extended this proposal and predicted a spatially inhomogeneous superconducting state which is presently known as FFLO state. This exotic imbalanced superconducting state is unique as it has a spatially-modulated order parameter, while the standard Bardeen-Cooper-Schrieffer(BCS) superconducting state has a spatially-homogeneous order parameter. The existence of the FFLO state is surprising in the sense that it retains superconductivity overcoming the orbital and Pauli-paramagnetic pair-breaking effects, even at very high magnetic fields. For this reasons the imbalanced systems has been studied vigorously - both 
theoretically and experimentally. Theoretical studies on imbalanced systems often focus on the possibility of exploring imbalanced superconducting states in different physical systems, for example, in population-imbalanced Ultracold atomic gases \cite{liao10,leo11,torma11,arghya2}, optical lattices\cite{cai11}, heavy-fermionic superconductor CeCoIn$_{5}$\cite{agterberg09},  two-dimensional organic superconductors\cite{croitoru12prb,croitoru12prl,arghya1,arghya3} and quark matter core of the neutron stars\cite{rajagopal01,casal04}. The experimental search is a topic of vigorous research till date as it is very hard to pinpoint this state in the phase diagram. In an experiment involving an imbalanced system, one can find the imbalanced state if: (i) the superconductor is in the clean limit and (ii) the value of Maki parameter is greater than 1.8. The most promising experimental systems in this context are the heavy-fermionic superconductor CeCoIn$_{5}$ \cite{young07,kenzelmann08,bianchi03,kumagai11} and quasi 
two-dimensional(2D), organic superconductors like $\kappa$-(BEDT-TTF)$_2$Cu(NCS)$_2$, in which BEDT-TTF is  bisethylenedithio-tetrathiafulvalene\cite{lortz07,beyer12}. So, even 50 years after its prediction, this field of imbalanced superconductivity remains an active field full of surprises ({\it For a review see} \cite{pita} {\it and references therein.}).       
   
\subsection{Holographic description}
More recently there has been a lot of effort \cite{Erdmenger11,Bigazzi11,pinz,Alsup12,Alsup1208,Alsup13} to understand the imbalanced superconducting systems using holography and AdS/CMT. Generally, the bulk gravitational Lagrangian which holographically describe an imbalanced superconductor is given by
\begin{eqnarray}
{\cal L} = \frac{\sqrt{-g}}{2k_4^2}\left(R + \frac{6}{L^2} - \frac{1}{4}F_{ab}F^{ab} - - \frac{1}{4}Y_{ab}Y^{ab} - V(|\phi|) - |\partial\phi - iqA\phi|^2 \right)
\label{lagr}
\end{eqnarray}
which is comprised of the AdS gravity with $\Lambda=-\frac{6}{L^2}$, two $U(1)$ gauge fields with field strengths
\begin{eqnarray}
F=dA,~~~~~Y=dB,
\label{fandp}
\end{eqnarray}
and one scalar field ($\phi$) with potential 
\begin{eqnarray}
V(|\phi|) = m^2\phi^{\dagger}\phi
\label{fandp}
\end{eqnarray}
which is charged under $U_{A}(1)$ but uncharged with respect to the other. 

As known from the AdS/CFT correspondence mass of the above bulk scalar field dictates the conformal dimension ($\Delta$) of the dual field in the following manner
\begin{equation}
\Delta(\Delta-3)=m^2L^2.
\end{equation} 
This relation is particularly helpful to capture the physics of an field theory operator with a conformal dimension of interest. For example to describe a Cooper pair type condensate which has $\Delta=2$, one fixes the mass of the bulk scalar field to be $m^2=-\frac{2}{L^2}$. Note that this choice does not violate the Brietelhoner-Freedman bound which for this case is $m^2\ge -\frac{9}{4L^2}$. Since our interest lies in this theoretical aspect, in this paper, we will fix the above mass value for the bulk scalar field in all our computations. For completeness it should be mentioned that other than mass, the scalar field also has a charge $q$, and different values of charge lead to different physical properties in the dual field theory.    

The above description of the gravitational system has the minimal ingredients needed to describe the superconductivity in the imbalanced systems. Starting from the equations of motion which include Einstein equations, Maxwell equations and a scalar field equation, one looks for the cases where the scalar field is zero and non-zero. The vanishing of the scalar field gives a normal Reissner-Nordstr{\"o}m black hole phase. On the other hand, if one finds a non-zero scalar field it is understood that a condensate has been formed in the dual field theory. Of course this situation has a serious contradiction with the black hole no-hair theorem that supports the vanishing scalar field, but the fact of getting non-zero scalar field in the context of holographic  superconductors hints that one needs to re-examine the no-hair theorem itself \cite{Hertog:2006rr,Gubser:2008px}. The above statement is true for any holographic superconductor. For the imbalanced case, with two $U(1)$ gauge fields with unequal chemical 
potential, we have the following additional advantage.

In imbalanced superconducting systems Cooper pair forms between two fermionic species with unequal chemical potentials (say $\mu_1$ and $\mu_2$). Now to capture this behavior in the dual gravitational theory, one needs two $U(1)$ bulk fields (say $U_A(1)$ and $U_B(1)$) with field strengths $A_a$ which accounts total chemical potential $2\mu=\mu_1+\mu_2$ and $B_{a}$ which accounts the mismatch $2\delta\mu(=\beta\mu)=\mu_1-\mu_2$ of those fermionic species in boundary theory.

With these preliminaries we now move to the next sections to deal with the equations of motion and to compute the HEE separately for black hole and superconducting phases.     

\section{Equations of motion}
Extemizing the Lagrangian (\ref{lagr}) with respect to various fields one has the following set of equations:

Einstein equation,
\begin{eqnarray}
G_{ab}+\frac{1}{2}\Lambda g_{ab} = -\frac{1}{2}T_{ab}
\label{ee}
\end{eqnarray}
where the energy-momentum tensor of the matter field is defined as $T_{ab}=\frac{2}{\sqrt{-g}}\frac{\delta {\cal L}_{matter}}{\delta g^{ab}}$.

 Maxwell equations for $A_a$ and $B_a$ fields reads
\begin{eqnarray}
\frac{1}{\sqrt{-g}}\partial_a(\sqrt{-g}g^{ab}g^{cd}F_{bc}) &=& iq g^{dc}[\phi^{\dagger}(\partial_c\phi - iqA_c\phi) - \phi(\partial_c\phi^{\dagger} + iqA_c\phi^{\dagger})] \label{meq1}\\
\frac{1}{\sqrt{-g}}\partial_a(\sqrt{-g}g^{ab}g^{cd}Y_{bc}) &=&0
\label{meq2}
\end{eqnarray} 
where the scalar/gauge coupling takes place only in $U_{A}(1)$ sector. 

In addition there is also a scalar field equation given by
\begin{eqnarray}
\frac{1}{\sqrt{-g}}\partial_a[\sqrt{-g}g^{ab}(\partial_b\phi - iqA_b\phi)] + iqg^{ab} A_b(\partial_a\phi - iqA_a\phi) + \frac{\phi}{2|\phi|}V'(|\phi|)=0
\end{eqnarray}

In order to proceed further we consider the follwing background metric
\begin{eqnarray}
ds^2=-g(r)e^{-\chi(r)} dt^2 + \frac{r^2}{L^2}(dx^2+dy^2) + \frac{dr^2}{g(r)}
\label{metr1}
\end{eqnarray}   
where $\chi(r)$ accounts for the backreaction due to matter fields. For a case where backreaction is negligible one sets $\chi=0$.  For all matter fields, the anstaz is assumed to be homogeneous
\begin{eqnarray}
\phi=\phi(r),~~~~~A_adx^a = \psi(r)dt, ~~~~~ B_adx^a=v(r) dt
\end{eqnarray}

Now one finally unwind all field equations by substituting the ansatz. The final set of equations now have two independent Einstein equations
\begin{eqnarray}
\frac{1}{2}\phi'^2+\frac{e^{\chi}(\psi'^2+v'^2)}{4g} + \frac{g'}{g r} + \frac{1}{r^2} - \frac{3}{gL^2} + \frac{V(\phi)}{2g} + \frac{e^{\chi}q^2 \phi^2 \psi^2}{2g^2} = 0 \label{ee1}\\
\chi' + r(\phi'^2 +  \frac{e^{\chi}q^2 \phi^2 \psi^2}{g^2}) = 0 \label{ee2}
\end{eqnarray}
two Maxwell equations for $\psi$ and $v$ fields
\begin{eqnarray}
\psi'' + \psi' (\frac{2}{r}+ \frac{\chi'}{2} ) - \frac{2q^2\phi^2}{g} \psi =0 \label{me1}\\
v'' + v'(\frac{2}{r}+ \frac{\chi'}{2} ) = 0 \label{me2}
\end{eqnarray}
and a scalar field equation
\begin{eqnarray}
\phi'' + \phi' (\frac{g'}{g} + \frac{2}{r} - \frac{\chi'}{2}) - \frac{V'(\phi)}{2g} + \frac{e^{\chi} q^2 \phi^2 \psi^2}{2g^2} = 0. \label{sfe}
\end{eqnarray}
In the remaining part of our work we will look for the simultaneous solution of the above set of equations to compute the HEE. From now on we set $2k_4^2=1,~L=1$.

\section{HEE for the normal (black hole) phase with varying $\beta$}

At high temperature (above $T_c$), when no superconductivity appears, one has a vanishing bulk scalar field. For such a case the right hand side of the Maxwell equation 
(\ref{meq1}) vanishes and the resulting solution of the set of field equations is a doubly charged Reissner-Nordstr{\"o}m black hole given by the metric
\begin{eqnarray}
ds^2 &=& -f(r)dt^2 + \frac{dr^2}{f(r)} + r^2(dx^2+dy^2),\label{rnads}\\
f(r) &=& r^2 (1-\frac{r_H^3}{r^3}) + \frac{\mu^2 r_H^2}{4r^2}(1-\frac{r}{r_H})(1+\beta^2) \\
\beta &=& \frac{\delta\mu}{\mu}
\end{eqnarray}
where the gauge fields are
\begin{eqnarray}
\psi(r)=\mu(1-\frac{r_H}{r})= \mu- \frac{\rho}{r} \label{gf1},\\
v(r) = \delta\mu (1-\frac{r_H}{r}) = \delta\mu - \frac{\delta\rho}{r}.\label{gf2}
\end{eqnarray}
Hawking temperature of this RN-AdS spacetime is given by
\begin{eqnarray}
T_{bh} &=& \frac{r_H}{16\pi}[12-\tilde{\mu}^2(1 + \beta^2)], \label{temp1}\\
\tilde{\mu} &=& \frac{\mu}{r_H}.
\end{eqnarray}

To compute HEE we change the radial coordinate from $r$ to $z=\frac{r_H}{r}$. This redefinition has some computational convenience. In the $t,~z,~x,~y$ system the metric 
looks like
\begin{eqnarray}
ds^2 &=& -r_H^2e^{-\chi}g(z)dt^2 + \frac{dz^2}{z^4g(z)} + \frac{r_H^2}{z^2} (dx^2 + dy^2), \label{rnads2}\\
g(z) &=& \frac{1}{z^2} - [1+\frac{{\tilde{\mu}}^2}{4}(1+\beta^2)]z +  z^2\frac{{\tilde{\mu}}^2}{4}(1+\beta^2). \label{fzrn}\\
\end{eqnarray}

The HEE expression (\ref{HEE}) now simplifies to 
\begin{eqnarray}
S_E &=& \frac{1}{4}\int_{0}^{L_y}\int_{-l/2}^{l/2}\sqrt{h}~dxdy \label{eern1}\\
       &=& \frac{L_y r_H}{4}\int_{-l/2}^{l/2}\frac{1}{z^2}\left(r_H^2 + \frac{z'^2}{z^2g(z)}\right)^{1/2}~dx \label{eern11} 
\end{eqnarray}
where `${h}$' is the determinant of the induced metric of the codimension 2 hypersurface and in the second equality prime denotes derivative with respect to $x$. Equation 
(\ref{eern11}) also tells us that the system is equivalent to one defined by the Lagrangian $L =\frac{1}{r_H z^2}\left(r_H^2 + \frac{z'^2}{z^2f(z)}\right)^{1/2}$.  In order to 
take into accout that the surface is minimal, we extremize the Lagrangian.  This extremization problem has a constant of motion which is nothing but the canonical Hamiltonian. 
In this way we obtain a measure of how the entangling surface is extended within the bulk (towards the horizon) and gives an infrared cut-off ($z_0$) on the integrating variable,
given by
\begin{eqnarray}
\frac{1}{z_0^2} = \frac{r_H}{z^2}\frac{1}{\sqrt{r_H^2+\frac{z'^2}{z^2g(z)}}} \label{com}
\end{eqnarray} 
Then converting the integrating variable from $x$ to $z$ the final expression of HEE reads as
\begin{eqnarray}
S_{E} &=& \frac{L_y r_H^2}{2} \int^{z_0}_{\epsilon} \frac{z_0^2}{z^3}\frac{1}{\sqrt{(z_0^4-z^4)g(z)}}\label{heern2}\\
 && = S_{f} + S_{div},
\end{eqnarray}
where $S_{f}$ and $S_{div}$ parts denote the finite and diverging part of the entanglement entropy as discussed earlier.

On the other hand the width of the subsystem `$A$' is expressed as 
\begin{eqnarray}
\frac{l}{2} &=& \int_{0}^{l/2} dx\\
&& =\frac{1}{r_H}\int^{z_0}_{\epsilon} \frac{zdz}{\sqrt{g(z)(z_0^4 - z^4)}}
\label{sizel}
\end{eqnarray}
So finally to explore the behavior of the HEE, one now needs to evaluate the expressions (\ref{heern2}) and (\ref{sizel}). For this we need to find the metric function $g(z)$ 
for different cases - namely for the AdS-RN black hole and the imbalanced superconductor, set the UV cut-off $\epsilon$ to a small value and consider $z_0$ near to the horizon. 
We note that by changing $z_0$ it is possible to study the behavior of $S_f$ as a function of the strip width $l$.

Doing this is easy for the black hole phase since we know the black hole metric explicitly. In Figure \ref{eeadsrn} we show the variation of HEE with respect to the strip
width for a fixed temperature $T_{bh}=0.13$ and for different values of the imbalance parameter $\beta$. The larger $l$ corresponds to the infra-red limit \cite{Albash:2012pd}. 
In addition to conforming the earlier results \cite{Albash:2012pd}, from this set of plots we find that if one keeps the system-size as well as temperature constant, 
HEE for RN-AdS phase increases with the increase in chemical potential imbalance $\beta$. This has the following important physical consequence: if one considers HEE as a measure of the number of degrees of freedom of a system, the plots in figure \ref{eeadsrn} tell us, that, for a system of given width and temperature, larger $\beta$ corresponds more degrees of freedom. 

Now we move to the next section where we examine the superconducting case. We shall approach the problem in a complete numerical set up. 

\begin{figure}
\centering
\includegraphics[scale=0.8]{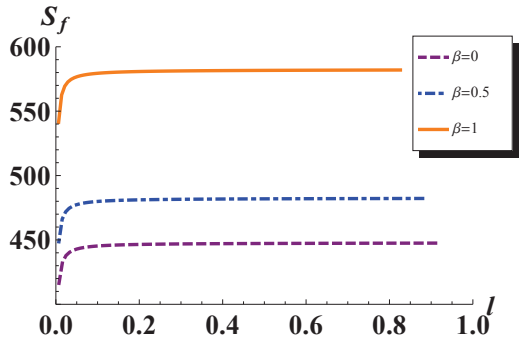}
\caption{(Color Online) Plot of holographic entanglement entropy as a function of the system's strip width $l$ for the AdS-RN black hole with different values of the 
imbalanced parameter $\beta.$ Here we have also set $\mu=1$ and temperature $T_{bh}=0.13$. Since $\mu$ and $\beta$ are fixed we adjust the horizon radius $r_H$ to keep the temperature constant.}
\label{eeadsrn}
\end{figure}

\section{HEE for the superconducting phase with varying $\beta$}

We now intend to calculate the HEE when the black hole has developed a scalar hair, in other sense, a superconducting state has been formed in the boundary field theory. 
It is not possible to compute the hairy black hole metric by staying within analytical limit. Therefore we approach this problem with the help of numerics. 

\subsection{Field equations and the bulk/boundary expansions}

We rewrite the equations of motion (\ref{ee1} to \ref{sfe}) by expressing $g(z)=\frac{r_H^2}{z^2}+h(z)$ which is helpful for further computations \cite{Bigazzi11}. 
In terms of $h(z)$ (and rescaled $\phi= z \phi$) these look like:
\begin{eqnarray}
&&\frac{\phi^{'2}}{2}+\frac{\phi\phi'}{z}+\frac{\phi^2}{2z^2}+\frac{e^{\chi}(\psi^{'2}+v^{'2})}{4(r_H^2+z^2h)}-\frac{h'}{z(r_H^2+z^2h)}+\frac{m^2r_H^2\phi^2}{2z^2(r_H^2+z^2h)}\nonumber\\
&+&\frac{1}{z^4}-\frac{r_H^2}{z^4(r_H^2+z^2h)}+\frac{e^{\chi}r_H^2q^2\psi^2\phi^2}{2(r_H^2+z^2h)^2}=0 \label{eez1}\\
&&\chi'-z\phi^2-\frac{z^3e^{\chi}r_H^2q^2\psi^2\phi^2}{(r_H^2+z^2h)^2}-2z^2\phi\phi'-z^3\phi^{'2}=0 \label{eez2}\\
&&\frac{\psi''}{r_H^2}+\frac{\psi'\chi'}{2r_H^2}-\frac{2q^2\psi\phi^2}{r_H^2+z^2h}=0 \label{mez1}\\
&&\frac{v''}{r_H^2}+\frac{v'\chi'}{2r_H^2}=0 \label{mez2}\\
&&\phi''+\left(\frac{2}{z}-\frac{2r_H^2}{z(r_H^2+z^2h)}-\frac{\chi'}{2}+\frac{h'z^2}{r_H^2+z^2h}\right)\phi' -\frac{r_H^2m^2\phi}{2z^2(r_H^2+z^2h)}\nonumber\\
&+&\left(-\frac{2r_H^2}{z^2(r_H^2+z^2h)}+\frac{q^2 e^\chi r_H^2 \psi^2}{(r_H^2+z^2h)^2}-\frac{\chi'}{2z}+\frac{h'z}{r_H^2+z^2h}\right)\phi =0. \label{sfez1}
\end{eqnarray}
In order to set the stage one needs to translate the problem of finding the hairy black hole into a boundary value problem by using Taylor series expansion of various fields. Near the horizon $z_H=1$ they are expanded as
\begin{eqnarray}
h_H(z)&=&-r_H^2+h_{H1}(1-z)+h_{H2}(1-z)^2+\cdots \label{hH}\\
\chi_H(z)&=&\chi_{H0}+\chi_{H1}(1-z)+\chi_{H2}(1-z)^2+\cdots\\
\psi_H(z)&=&\psi_{H1}(1-z)+\psi_{H2}(1-z)^2+\cdots\\
v_H(z)&=&v_{H1}(1-z)+v_{H2}(1-z)^2+\cdots\\
\phi_H(z)&=&\phi_{H0}+\phi_{H1}(1-z)+\phi_{H2}(1-z)^2+\cdots \label{phihor}
\end{eqnarray}
Note that in the Taylor expansion of $h_H(z)$, we set the first term as $-r_H^2$ to fulfill the requirement that the metric coefficient $g(z)$ vanishes at the horizon. Also, in order to prevent the gauge fields from acquiring infinite norm at the horizon one needs $\psi_H(z=1)=0=v_H(z=1)$. Therefore, upto a second-order expansion, one has twelve unknown coefficients in the Taylor expansions. However not all of them are independent, they are related by five equations (\ref{eez1} to \ref{sfez1}) and one needs to substitute the field expansions in these equations. This gives a set of algebraic equations which relate various Taylor coefficients in different orders of expansion. Finally one is left with six independent coefficients and all others are expressible in terms of them. We choose these coefficients to be $\phi_{H0}, \chi_{H0}, \psi_{H1}, v_{H1}, q, r_{H}$. The next step is to find the expressions of dependent Taylor coefficients appearing in the near horizon expansions in terms of these independent parameters. Some of them with relatively simpler expressions are:
\begin{eqnarray}
&& h_{H1}=-\frac{1}{4}e^{\chi_{H0}}(v_{H1}^2+\psi_{H1}^2)+r_H^2(1+\phi_{H0}^2)\\
&&\chi_{H1}=-\frac{16r_H^2(r_H^2+e^{\chi_{H0}}q^2 \psi_{H1}^2)\phi_{H0}^2}{(e^{\chi_{H0}}(v_{H1}^2+\psi_{H1}^2)-4r_H^2(3+\phi_{H0}^2))^2}\\
&& \phi_{H1}=\phi_{H0}+\frac{4r_H^2\phi_{H0}}{e^{\chi_{H0}}(v_{H1}^2+\psi_{H1}^2)-4r_H^2(3+\phi_{H0}^2)}\\
&& \psi_{H2}=\frac{4r_H^2\psi_{H1}\phi_{H0}^2 \left(-e^{\chi_{H0}} q^2 v_{H1}^2+r_H^2 \left(1+4 q^2 \left(3+\phi_{H0}^2\right)\right)\right)}{\left(e^{\chi_{H0}} \left(v_{H1}^2+\psi_{H1}^2\right)- 4 r_H^2
\left(3 + \phi_{H0}^2\right)\right)^2}\\
&& v_{H2} = \frac{4r_H^2 v_{H1}\left(e^{\chi_{H0}} q^2 \psi_{H1}^2+r_H^2\right)\phi_{H0}^2}{\left(12r_H^2-e^{\chi_{H0}} v_{H1}^2-e^{\chi_{H0}} \psi_{H1}^2+4r_H^2\phi_{H0}^2\right)^2}.
\end{eqnarray}
The others are more complicated having a large number of terms. We prefer not to write them here explicitly.

Now let us write down the ultraviolet (UV) asymptotic (boundary) behavior of all fields near the AdS boundary $z=0$:
\begin{eqnarray}
h_b(z)&=& -\frac{\epsilon}{2r_H}z+\cdots\\
\chi_b(z)&=& \log (1+a) = 0\\
\psi_b(z)&=& \mu - \frac{\rho~ z}{r_H}+\cdots\\
v_b(z)&=& \delta\mu - \frac{\delta\rho~ z}{r_H}+\cdots\\
\phi_b(z)&=& \frac{C_1}{r_H} + \frac{C_2}{r_H^2} z  +\cdots
\end{eqnarray}
where $\epsilon$ is the mass of RN-AdS black hole defined at the spatial asymptote. As usual, both $C_1$ and $C_2$ cannot be nonzero at the same time. Here our aim is to solve the 
boundary value problem with $C_1=0$ but $C_2\ne0$. The reason behind this is that $C_2$ has conformal mass dimension 2 which corresponds to $\Delta=2$ of the Fermionic operator representing the condensate.

\subsection{Numerical scheme for finding the hairy black hole}
Here we look for the solution of the above set of equations in order to compute the hairy black hole metric. Our focus thus is on getting the solution for $h(z)$. We use the shooting method for this purpose. Here the basic idea for solving the boundary value problem is to first express various boundary parameters in terms of near horizon fields and their derivatives. For that one inverts the above set of equations to write
\begin{eqnarray}
\mu &=& \psi_b (z) - z \psi'_b (z), \label{mu} \\
\rho &=& -r_H \psi'_{b} (z), \label{rho} \\
C_{1} &=& \phi_{b} (z) r_{H} - \phi'_{b}(z) r_H z, \label{c1} \\
C_{2} &=& \phi'_{b} (z) r_{H}^{2}, \label{c2} \\
a &=&  e^{-\chi_b (z)}-1, \label{a} \\
\epsilon &=& -2r_{H} h'_{b} (z), \label{m} \\
\delta\rho &=& - r_H v'_{b} (z),\label{drho}\\
\delta\mu &=& v_{b} (z) - z v'_{b} (z). \label{dmu}
\end{eqnarray}
The temperature of this superconducting state is also given in terms of near horizon expressions, given by \cite{Bigazzi11}
\begin{eqnarray}
T_{sc}=  \frac{r_H}{16\pi}\left((12+4\phi_{H0}^2)e^{-\frac{\chi_{H0}}{2}} - \frac{1}{r_H^2}e^{\frac{\chi_{H0}}{2}}(\psi_{H1}^2 + v_{H1}^2)\right).
\label{tempsc}
\end{eqnarray}
The critical temperature corresponds to the smallest possible value of $\phi_{H0}$ such that the hair is just developed. 

In the numerical scheme we set a very small value for $\phi_{H0}$ and all other parameters are fixed by hand and they are provided as the input {\it seed} to solve the set of coupled differential equations. Then in the following step we make a very small increment for the seed of $\phi_{H0}$ and that will determine other near horizon parameters which, together, will set the input values for the second step. Moreover, at every step, one finds the values of various UV parameters (\ref{mu} to \ref{dmu}) as a part of the output. In this way one generates a set of data of solutions by implementing this iteration for a number of times. For each iteration one has numerical values for--(i) near-horizon parameters and (ii) boundary parameters as a function of that. It is then trivial to reproduce $h(z)$ as well as the metric $g(z)$. Furthermore at each step we get a temperature given by Eq. (\ref{tempsc}). As we mentioned earlier that our aim is to find the hairy black hole solution so that we can use that for the further computation of the HEE. 

In Figure 3 we plot the family of hairy black hole metrics $g(z)$ for two cases with $\beta=0.01$ and $\beta=0.02$. For a fixed imbalanced parameter, different plots correspond to distinct temperatures which in our case are very close to each other. Subsequently we shall choose one of these metrics with a particular temperature and compute the HEE to compare with the black hole phase.

Before going further some words about our code are in order. As usual, while solving the boundary value problem the issues with divergences are tackled by propagating the near horizon solutions from $\epsilon_H = 0.00001$ away from the horizon ($z=1$) to $\epsilon_b=0.000001$ near the boundary ($z=0$). As the boundary conditions we have set $C_1=0, a=0, \mu=1$ and $\delta\mu=0.01$ for one case while $\delta\mu=0.02$ for the other. In all cases we have set $q=2$.

\begin{figure}
\centering
\includegraphics[scale=0.6]{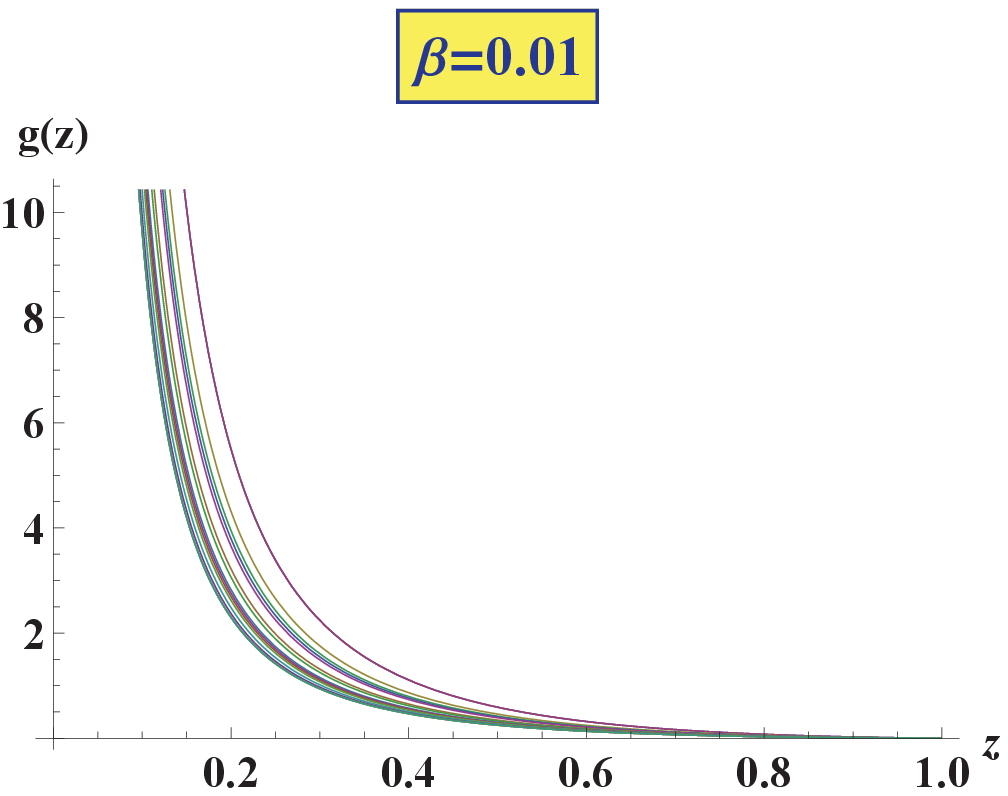}
\includegraphics[scale=0.7]{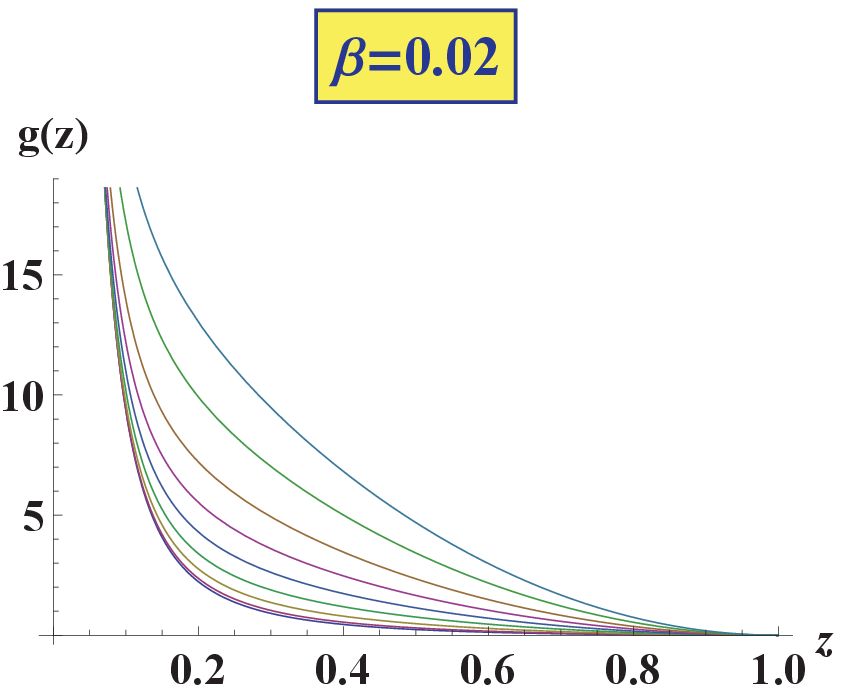}
\caption{(Color Online) Plot of the hairy black hole metric function with respect to the radial coordinate $z$ as obtained from numerical computations. As expected $g(z)$ vanishes at the horizon and diverges at asymptotic infinity. Different curves in this plot correspond to different characteristic temperature.}
\label{eehs}
\end{figure} 
Now we move to the final part where we calculate the HEE for the superconducting phase and compare with the black hole phase.

\subsection{HEE for the superconducting phase and comparison with the black hole phase}
Having found the metric functions $g(z)$ and corresponding temperatures we are free to choose one entry of this set and use Eq.(\ref{heern2}) and Eq. (\ref{sizel}) with varying $z_0$ to obtain the list required for plotting $S_{f}$ as a function of $l$. In order to compare with the AdS-RN black hole phase we fix the black hole temperature $T_{bh}$ (\ref{temp1}) to be equal to the preassigned temperature for the superconducting phase (given by Eq. \ref{tempsc}) by suitably adjusting its horizon radius (since $\mu$ and $\delta\mu$ are already fixed). With this new horizon radius then we use the black hole metric to calculate the list for the required plot. 

\begin{figure}
\centering
\includegraphics[scale=0.6]{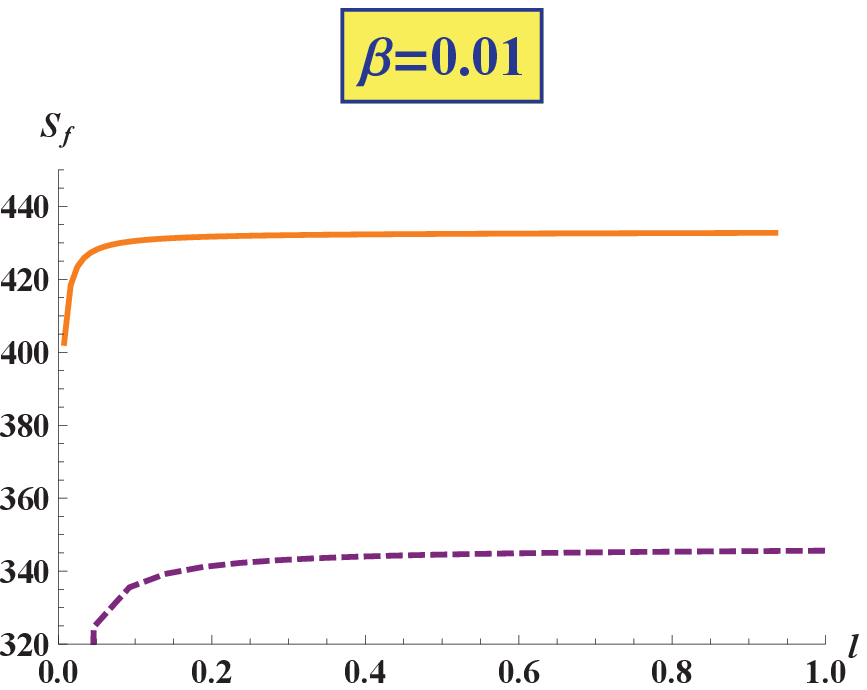}
\includegraphics[scale=0.6]{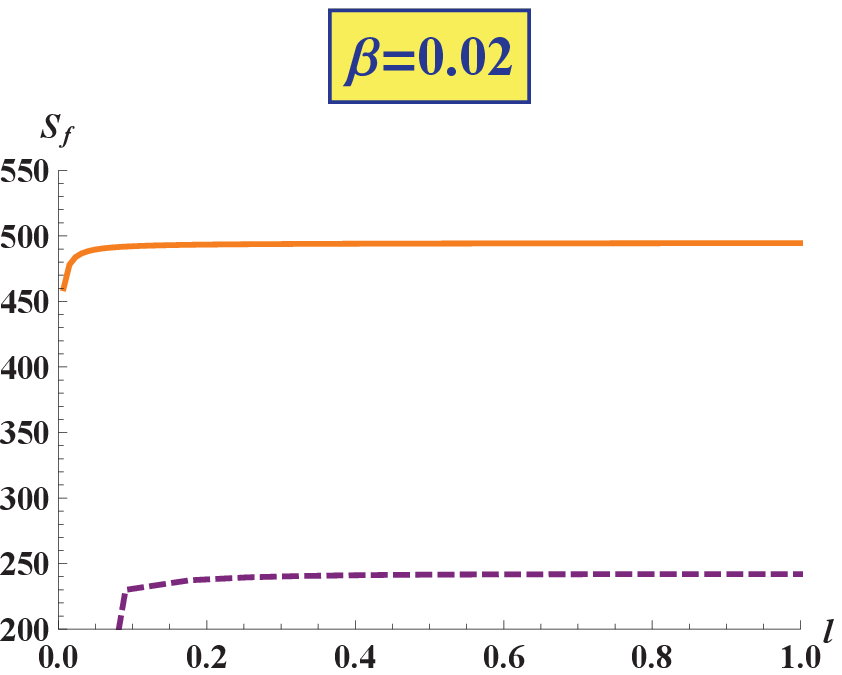}
\caption{(Color Online) Plot of holographic entanglement entropy of AdS-RN black hole (solid lines) and imbalanced superconductors (dashed lines) at $T_{bh}=T_{sc}=0.13$ for $\beta=0.01$ and $\beta=0.02$. The results remain similar if we choose any other temperature and $g(z)$ from Figure 3. For other details refer to text.}
\label{compare1}
\end{figure}

In Figure \ref{compare1} we compare the relative values of the HEE between the black hole and superconducting states for fixed (i) chemical potential $\mu =1$, (ii) imbalance parameter $\beta=0.01,~0.02$ respectively and (iii) identical values of temperatures $T_{bh}=T_{sc}$ for each $\beta$. From both of these plots we note that the superconducting state has a lower HEE than the normal (RN-AdS) state. This, as explained by Albash et al\cite{Albash:2012pd}, represents the fact that the degrees of freedom have condensed from the RN-AdS case to the superconducting state and may serve the purpose of signaling the preferable state.

\subsection{Variation of HEE with $\beta$ for the holographic superconductor}
Finally we are in a position to compare the change in HEE for the superconducting phase for different imbalances while all other parameters are kept fixed. For this we do not need to perform anymore computations, rather, we compare the superconducting phase plots from Figure 4. This is depicted in Figure 5 which shows that with the increase in chemical imbalance HEE decreases. Notably this behavior is exactly opposite to the RN-AdS phase as shown in Figure 2. 

In order to understand this difference physically one should consider the fact that thermodynamics of AdS black holes may differ substantially from a physical system like superconductors. For example if we, keeping the horizon radius constant, increase $\beta$ then temperature of the RN-AdS black hole as given by Eq. (\ref{temp1}) becomes smaller. On the other hand one can check that HEE increases with the increase $\beta$ for constant horizon radius. Since HEE in certain cases resembles with the black hole entropy \cite{Takayanagi:2012kg} one can roughly interpret this behavior in terms of the negative specific heat of black holes. It is known that in certain cases AdS black holes do have negative specific heat \cite{Banerjee:2011raa}. On the other hand one expects a superconducting system to have a positive heat capacity and therefore the difference with the black hole phase is natural.

On the other hand this behavior might challenge the HEE to correctly identify the preferable state. The fact that HEE for black hole phase increases while it decreases for the superconducting state implies that for larger chemical potential imbalance superconducting state will be more probable. Of course this goes against the fact that with arbitrarily large imbalance one cannot achieve superconductivity. Therefore one should be careful in interpreting physics of holographic superconductors only by looking at the HEE.

\begin{figure}
\centering
\includegraphics[scale=0.7]{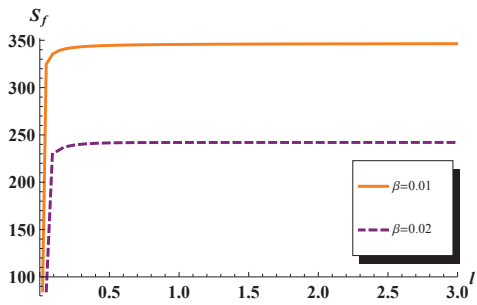}
\caption{(Color Online) Plot of holographic entanglement entropy of the superconducting phase for two different chemical imbalances. In these plots $\mu=1$ and $T_{sc}=0.13$ for both cases. The results remain similar if we choose any other temperature and $g(z)$ from Figure 3. Details are given in text.}
\label{compare}
\end{figure}

\section{Conclusions and Discussions}
In this paper we computed Holographic entanglement entropy (HEE) starting from a fully back-reacted gravitational theory which describes imbalanced superconductivity below the critical temperature and and doubly-charged RN-AdS black hole at temperature higher than the critical temperature. We chose the strip geometry for the entangled surface and compute the HEE as a function of strip size. The hairy black hole metric was found by using the numerical shooting method. Results showed that HEE for the superconducting state is lower than the black hole/normal phase for the values of the imbalance parameter ($\beta$) considered in this work. It was also shown that the effect of the imbalance is exactly opposite for black hole and superconducting phases. For the AdS-RN black hole phase HEE increases with the increase in the imbalance in two chemical potentials. Whereas for holographic superconductor HEE decreases. The fact that HEE for imbalanced holographic superconductor (also for other cases reported earlier \cite{Albash:2012pd}-\cite{Cai:2013oma}) is less than the black hole might insist one to consider this as a good physical parameter to identify the preferable state below $T_c$. 

The present study also raises a question whether or not HEE \textit{alone} can always correctly identify the preferable state for physical systems like imbalanced mixtures. The fact that HEE increases for the black hole phase and decreases for superconducting phase with respect to increasing imbalance implies that superconducting state will be more and more preferable as imbalance increases. But, as known from physical considerations, this is not the case with imbalanced systems. So clearly HEE fails to serve this purpose in this context. Usually for a condensed matter system one uses free energy in order to say anything about the preferable state. With the concern we mentioned it is unlikely that HEE alone could serve the purpose of free energy for holographic superconductors.

\section*{ Acknowledgement:}
S.K.M thanks Saurav Samanta for useful discussions in the early stages of this work. A.D thanks IUCAA, Pune for kind hospitality and support during his visit. He also thanks Council of Scientific and Industrial Research, India for financial support in the form of fellowship (File No.09/575(0062)/2009-EMR-1).

\end{document}